\documentclass[reprint,aps,prb,showpacs,amsmath,superscriptaddress]{revtex4-1}
\usepackage{bm}
\usepackage{graphicx}
\usepackage{color}
\usepackage{amssymb}

\newcommand{\FeZr}{Fe$_{90}$Zr$_{10}$}
\newcommand{\AlZr}{Al$_{75}$Zr$_{25}$}
\newcommand{\Tc}{T_{\mathrm{c}}}

\begin{document}

\title{Finite size effects in amorphous {\FeZr/\AlZr} multilayers}

\author{P. T. Korelis}
\affiliation{Department of Physics and Astronomy, Uppsala
University, Box 516, SE-751 20 Uppsala, Sweden}
\author{P. E. J{\"o}nsson}
\affiliation{Department of Physics and Astronomy, Uppsala
University, Box 516, SE-751 20 Uppsala, Sweden}
\author{A. Liebig}
\affiliation{Department of Physics and Astronomy, Uppsala
University, Box 516, SE-751 20 Uppsala, Sweden}
\affiliation{Institute of Physics, Chemnitz University 
of Technology, D-09107 Chemnitz, Germany}
\author{H.-E. Wannberg}
\affiliation{Department of Physics and Astronomy, Uppsala
University, Box 516, SE-751 20 Uppsala, Sweden}
\author{P. Nordblad}
\affiliation{Department of Engineering Sciences, Uppsala
University, Box 534, SE-751 21 Uppsala, Sweden}
\author{B. Hj{\"o}rvarsson}
\affiliation{Department of Physics and Astronomy, Uppsala
University, Box 516, SE-751 20 Uppsala, Sweden}

\date{\today}

\begin{abstract}
The thickness dependence of the magnetic properties of amorphous  {\FeZr} layers has been explored using {\FeZr/\AlZr} multilayers.  The  {\AlZr} layer thickness is kept at 40~{\AA}, while the thickness  of the {\FeZr} layers is varied between 5 and 20~\AA. The thickness of the {\AlZr} layers is sufficiently large to suppress any significant interlayer coupling. Both the Curie temperature and the spontaneous magnetization decrease non-linearly with decreasing thickness of the {\FeZr} layers.  
No ferromagnetic order is observed in the multilayer with 5~{\AA}  {\FeZr} layers.
The variation of the Curie temperature $\Tc$ with the {\FeZr} layer thickness $t$  is fitted with a finite-size scaling formula  $[1-\Tc(t)/\Tc(\infty)]=[(t-t')/t_0]^{-\lambda}$, yielding $\lambda=1.2$, and a critical thickness $t'=6.5$~\AA, below which the Curie temperature is zero.
\end{abstract}


\maketitle

\section{Introduction}

The reduction of one of the spatial extensions of a material is accompanied by
large changes in magnetic properties. \cite{Bland94B, Freeman92R}  
The change of the surface/volume ratio affects 
the magnetic anisotropy energy\cite{Poulop99R, Bader94} and the presence of proximity effects can lead to an effective enhancement 
or reduction of the magnetic moment. \cite{Vaz08R} 
Finite-size scaling theory \cite{Allan70,Fisher72}  predicts a reduction of the transition temperature as the film thickness is reduced to the nanometer scale. This is experimentally observed for e.g. the Curie temperature of a number of crystalline thin film systems,\cite{Vaz08R,Li92,Huang93} as well as for the spin glass transition temperature of thin spin-glass layers. \cite{Kenning87}
A finite-size scaling of the form $[1-\Tc(t)/\Tc(\infty)]=(t/t_0)^{-\lambda}$, where the shift in transition temperature $\Tc(t)$ with changing film thickness $t$ is given by the shift exponent $\lambda=1/\nu$, is usually used to fit experimental data. This relation has been shown to hold experimentally for films in the 3D limit, but not in the thickness limit where a 3D to 2D crossover occurs.\cite{Li92}
An alternative scaling relation was proposed for films in the ultrathin film limit\cite{Allan70,Huang93}
\begin{equation}
1-\Tc(t)/\Tc(\infty)=[(t-t')/t_0]^{-\lambda}
\label{eq:FSS}
\end{equation}
where $\Tc(t)=0$ at a finite film thickness $t'$.
This scaling relation has been shown to hold even through a 3D-2D crossover. \cite{Huang93}

The magnetic properties of crystalline and 
amorphous materials are similar in some respect, but there are well defined and  important differences, such as the presence of local random anisotropy \cite{Thomas92, Suzuki08} instead of magnetocrystalline anisotropy and the absence of atomic steps, and hence step-edge anisotropy,  at interfaces of amorphous layers.

 An amorphous Fe layer can be interface-stabilized, in certain multilayer structures, up to a critical thickness of the amorphous layer, \cite{Castano99,Dubowik95,Pan99,Kopcewicz97,Handschuh93,Honda94,Thiele93,Landes91}
above which the bcc structure of Fe is recovered. 
Interface stabilization has therefore limited value when exploring the thickness 
dependence of the magnetic properties of amorphous Fe, both due to the limits in the accessible range and the possible changes in the atomic 
configurations with the layer thickness.
An alternative route for obtaining amorphous layers is to form alloys of elements with large differences in atomic size. 
Zr can be used to make amorphous Fe-based alloys, in which a minute amount of Zr can cause a large increase of the critical thickness for the amorphous-to-crystalline transition.
At concentrations above 7~at.~\%, the amorphous phase is fully stabilized and amorphous FeZr alloys can therefore be deposited without restrictions in thickness. \cite{Mayr02, Geisler96}

There exists a wealth of reports on the bulk properties of amorphous FeZr alloys from both 
 experimental \cite{Hiroyoshi82,Fukamichi82,Shirakawa83,Read84,Saito86,Ryan87,Tange89,Kaul91,Kaul92B,Kiss94,Ren95,Wildes03,Calderon05} 
and theoretical studies. \cite{Lorenz95,Uchida01,Park10}
Currently there is no generally accepted view on the description of the magnetic structure at the atomic-scale, although several 
models have been proposed. These include the 'wandering axis' model, in which the axis of the local ferromagnetic order 
changes direction over short distances, \cite{Ryan87} the freezing of the transverse components of the magnetization at low 
temperature, \cite{Ren95} the formation of antiferromagnetic spin clusters in a ferromagnetic matrix 
\cite{Hiroyoshi82,Read84,Saito86} and the arrangement of the moments in finite, interacting, non-collinear spin 
clusters \cite{Kiss94,Wildes03} or non-collinear clusters embedded in an infinite ferromagnetic 
matrix. \cite{Kaul91,Kaul92B,Calderon05,Kaul05REV}

For amorphous FeZr/AlZr multilayers it has been experimentally observed that the interlayer exchange coupling between 
the magnetic layers can be neglected when the thickness of the AlZr layers exceeds 25~{\AA}. \cite{Liebig11}
Multilayers with several repetitions of the magnetic layer can therefore be used to explore inherent finite size effects. 
Here, we address the dependence of the {\FeZr} layer thickness on the magnetic ordering in amorphous {\FeZr/\AlZr} multilayers.
First, the structural quality of the  multilayers is established, and thereafter the resulting magnetic properties 
are presented. Finally, the effect of finite size on the Curie temperature ($\Tc$) and the size of the magnetic moment are discussed.

\section{Experimental details}

The samples were deposited at room temperature in a UHV magnetron sputtering system. The base pressure of the system was 
in the 10$^{-8}$ Pa range and the sputtering gas was high purity Ar,  at a pressure of 0.4 Pa. The substrates were 
Si (111) precut crystals with a thermal oxide surface layer. The sample stage was rotated during deposition, to ensure 
lateral uniformity in layer composition and thickness. The \FeZr~and \AlZr~alloy layers were co-deposited from three 
different DC sources, using high-purity targets. The desired composition was achieved by adjusting the sputter rate of 
each material. Rutherford Backscattering Spectrometry (RBS) measurements were performed to determine the composition of the 
layers.

Four different multilayer samples were investigated,  all prepared with a seed layer of \AlZr, upon 
which ten repetitions of the bilayer were deposited. The bilayer consists of an {\FeZr} layer of different thickness, 
varying from 5~{\AA} to 20~{\AA}, 
and an {\AlZr} layer with a nominal 
thickness of 40~{\AA}. Both the top and bottom interfaces for all {\FeZr} layers are \AlZr. A 300~{\AA} thick film 
of {\FeZr} on an {\AlZr} seed layer was prepared as reference for the bulk limit. A 25~{\AA} thick capping layer of Al was 
deposited on top of all samples. Each sample will be referred to by its nominal layer thickness, e.g. the sample with 
5~{\AA} {\FeZr} and 40~{\AA} {\AlZr} is labeled 5/40.

The layering and the amorphous quality of the samples were investigated by X-ray reflectivity and diffraction measurements. 
All reflectivity scans were performed using Cu K\textit{$_a$}
radiation on a Siemens D5000 diffractometer in Bragg-Brentano geometry with a secondary monochromator.
Diffraction scans with small incidence angle (grazing incidence) of $1.2^{\circ}$ with respect to the sample surface were 
carried out on a parallel beam setup (Cu K\textit{$_a$} radiation) equipped with a Goebel mirror and a Soller slit with $0.40^{\circ}$ divergence.
The ac-susceptibility and dc-magnetization measurements were performed in a Quantum Design MPMS SQUID system. Before each 
measurement series the remanent field of the superconducting magnet was nulled out using the ultra low field option of the 
MPMS system.

\section{Results and discussion}

\subsection{Structural properties}

\begin{figure}[th]
\centering
\includegraphics[width=7.5cm]{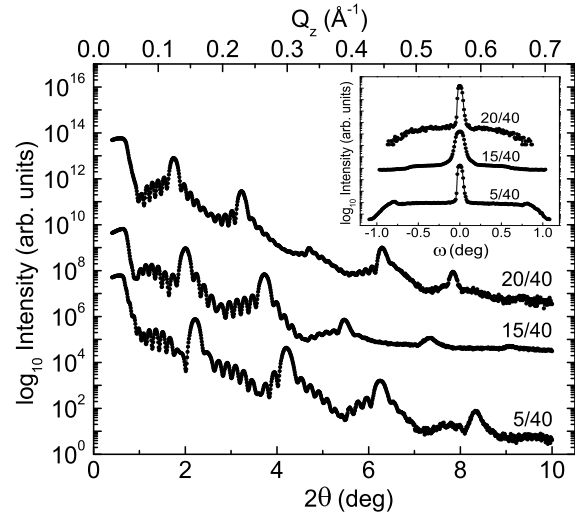}
\caption{
X-ray reflectivity scans for the multilayer samples with 5, 15 and 20~{\AA} thick {\FeZr} layers. 
The scans are offset along the intensity axis. 
Transverse scans on the first order bilayer peaks are shown in the inset. 
The positions of the observed maxima are set to zero.
\label{reflectivity}
}
\end{figure}

The X-ray reflectivity (XRR) curves for the samples with the thickest and thinnest \FeZr~layers (20/40 and 5/40, 
respectively) as well as the 15/40 sample are presented in Fig.~\ref{reflectivity}. For the 20/40 sample, bilayer peaks 
up to the fifth order are observed. The appearance of many reflectivity peaks in an extended $2\theta$ range is 
evidence of well-defined bilayer thickness with small interface roughness. 
The short period oscillations (Kiessig fringes) are visible in between the reflectivity peaks, signifying a well-defined total film thickness and small surface roughness. 
The results for all samples are fairly similar, except those obtained from the 15/40 sample.  A more rapid decrease of the reflected intensity is seen in the results from the 15/40 sample, with the fourth order peak only barely seen. The intensity of the Kiessig fringes also decays faster than for the 
other samples, indicating larger waviness in the structure.

We obtained the thickness of the individual layers by analyzing the reflectivity scans.
The results are summarized in table~\ref{thicknesstable}. 
The seed layer was determined to be 94(1)~{\AA} thick for all samples and the capping layer is between 25 and 30~{\AA}, depending on the 
degree of oxidation of the Al capping.

\begin{table}[hb]
\begin{center}
\begin{tabular}{|c| c| c|}
\hline
	& \multicolumn{2}{c|}{Layer thickness ({\AA})}\\
		\cline{2-3}
Sample & \FeZr & \AlZr \\
\hline
5/40 & 4.7(1) & 37.5(4)\\
10/40 & 9.5(1) & 37.5(4)\\
15/40 & 14.2(3) & 34.0(4)\\
20/40 & 18.9(1) & 37.5(4)\\
300 & 280(2) & - \\
\hline
\end{tabular}
\caption{The actual layer thickness for the samples, obtained from the X-ray reflectivity measurements. 
The uncertainty for the last significant digit is shown in parenthesis.}
\label{thicknesstable}
\end{center}
\end{table}

Transverse scans were measured in the vicinity of all the reflectivity peaks and the results  for the first order reflectivity peaks are shown in the inset of Fig.~\ref{reflectivity}. 
A narrow and sharp intensity profile near the specular condition and a broad background that corresponds to diffuse 
scattering can be seen. No significant changes are observed in the transverse scans when the {\FeZr} layer thickness is 
varied, except for the 15/40 sample which has considerably larger width and significantly higher diffuse scattering. The integrated intensity of the diffuse scattering in this sample adds up to 2.5~\% of the integrated 
specular intensity, while the corresponding results for the other multilayer samples are  0.5--1~\%. The 15/40 sample exhibits therefore less flat layers, which can be described as waviness in the layered structure.  

\begin{figure}[th]
\includegraphics[width=7.5cm]{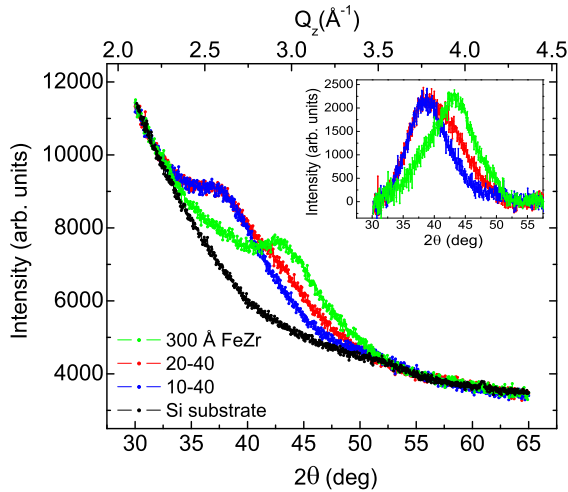}
\centering
\caption{
(Color online) Grazing incidence X-ray diffraction measurements for the 300~{\AA} {\FeZr} and multilayer samples. 
The inset shows the results with the substrate background subtracted. 
\label{gidiff}
}
\end{figure}

We performed grazing incidence X-ray diffraction (GIXRD) scans to investigate the amorphous quality of the samples. 
The GIXRD patterns for the bulk reference {\FeZr} layer and the multilayer samples are shown in Fig.~\ref{gidiff}. 
Only very broad features can be observed, which signifies a short coherence length of the interatomic distances and is 
characteristic of amorphous structures. The non-linear background originates from the top SiO$_{2}$ layer of the substrate.
The inset of Fig.~\ref{gidiff} shows the diffraction patterns with the background subtracted. 
The only detected peak for the 300~{\AA} {\FeZr} layer is centered at $43.6^{\circ}$.  For crystalline bcc Fe, the strongest diffraction line is found  at $44.7^{\circ}$ and corresponds to the distance between the $(110)$ atomic planes. Therefore, the peak position for the amorphous {\FeZr} layer indicates a slight increase of 
the average nearest neighbor distance for the Fe atoms.
For the multilayers, a broad peak centered at approximately $38.1^{\circ}$ is detected. This position is in the vicinity 
of the $(111)$ reflection from elemental fcc Al ($38.5^{\circ}$). The contribution to the multilayer diffraction patterns
from the {\FeZr} layers does not give rise to a distinguishable peak but is seen as an extended shoulder in the high-angle 
side.

\subsection{Magnetic properties}

\begin{figure}[h]
\includegraphics[width=7.5cm]{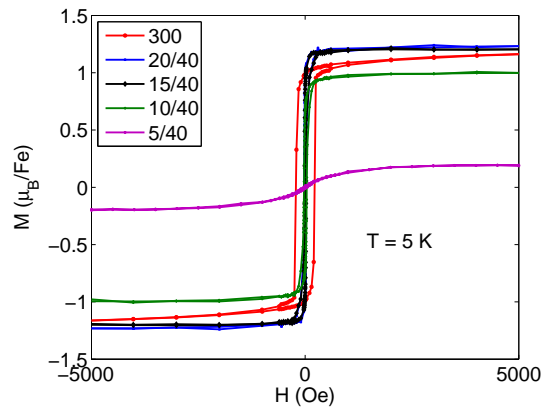}
\centering
\caption{
(Color online) Field dependence of the magnetization  recorded at $5$~K. 
A density for amorphous {\FeZr} of 8.0~g/cm$^3$ is used to obtain the magnetic moment per iron atom.
The data were corrected for the diamagnetic contribution of the Si-substrates.
\label{MH}
}
\end{figure}

The field dependence of the magnetization of the samples measured at 5~K is displayed in Fig.~\ref{MH}. All samples, except the 5/40 sample, show a ferromagnetic response.  
As seen in the figure, the $M(H)$ curves look similar for the ferromagnetic {\FeZr} multilayers, with small coercivity 
and low saturation fields. The 300~{\AA} film has larger coercivity, around 200~Oe, and shows a gradual increase of the 
magnetization with the applied field, which extends well above the apparent saturation field value. This behavior, commonly observed in amorphous FeZr bulk samples, can be 
understood as originating from alignment of regions with non-collinear components of the magnetic moments. The magnetic moment per iron atom is found to be 1.3~$\mu_B$/Fe in a field of 4~T for the 300~{\AA} sample. This value agrees well with other studies of  co-sputtered bulk-like films of \FeZr, \cite{Fukamichi82} and the magnetization value at 0.1~T agrees with the moment  obtained by polarized neutron reflectivity experiments (PNR) on the same type of thick {\FeZr} film.\cite{Korelis10}

\begin{figure*}[!t]
\includegraphics[width=15cm]{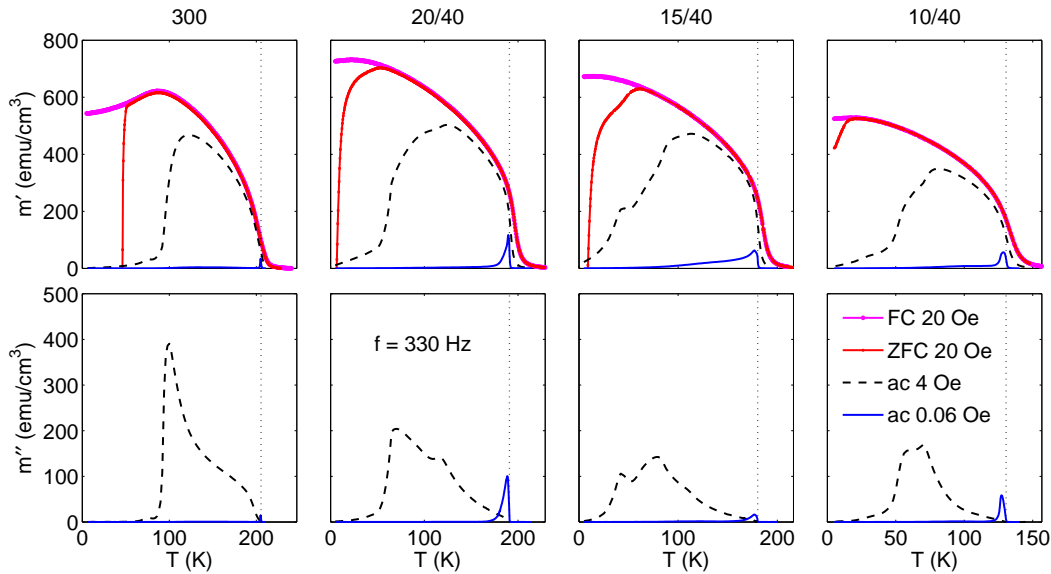}
\centering
\caption{
(Color online) Upper panels: Temperature dependence of the zero-field-cooled (ZFC), field-cooled (FC) and the real components  of 330~Hz ac magnetization. Lower panels: Temperature dependence of the imaginary components of 330~Hz ac magnetization.
The rms amplitudes of the oscillatory fields are 4 and 0.06~Oe. 
During the ZFC process, a small negative remanent field in the magnet gave rise to an initially negative ZFC 
magnetization for the 300~{\AA}, 20/40 and 15/40 samples. 
The Curie temperature (defined in the text) is indicated as a dotted line.
\label{MvsT-ac}
}
\end{figure*}

The results from zero-field-cooled (ZFC) and field-cooled (FC) magnetization measurements in a magnetic field of 20~Oe are displayed in Fig.~\ref{MvsT-ac}. Also shown in the figure are the real and imaginary components of the ac magnetization, measured with field amplitudes of 0.06 and 4.0~Oe at a frequency of 330~Hz. The transition temperature $\Tc$ of each sample was determined from the inflection point in the low field $m'(T)$ curve, and it is 
marked with a dotted line in the figure. The acquired $\Tc$ for the reference sample is 205~K, which agrees well 
with some earlier studies \cite{Saito86,Tange89} but is somewhat lower than other reported values for bulk, rapid-quenched 
alloys of the same nominal composition. \cite{Hiroyoshi82,Shirakawa83,Read84,Ryan87,Kiss94,Ren95,Wildes03}
The transition temperature of the multilayers decreases with decreasing magnetic layer thickness - being 191, 180 and 
130~K for the 20/40, 15/40 and 10/40 multilayers, respectively. 
For all samples, the $\Tc$ is found to coincide with the onset of the imaginary 
part of the susceptibility, $m''(T)$ (lower part of Fig.~\ref{MvsT-ac}).

Below $\Tc$, the FC and ZFC curves bifurcate and the bifurcation temperature is both sample- and field-dependent. The bifurcation temperature 
is governed by the temperature dependence of the coercivity and occurs at a temperature where the coercivity has reached the order of the applied field.  The observation of a splitting between the FC and ZFC curves does not prove or disapprove the existence of a possible 
re-entrant spin glass phase \cite{Hiroyoshi82,Saito86,Kaul92B} or transverse spin-freezing temperature. \cite{Ryan87,Ren95} 
However, the low field (0.06 Oe) ac magnetization results do not indicate any low-temperature anomaly characteristic of a glass transition for any of the samples.

The field dependence of the ac magnetization, shows significant difference in-between the reference  300~{\AA} {\FeZr} sample and the multilayers and also among the three multilayers. 
The very weak response of the reference sample to a small ac magnetic field of 0.06 Oe (discernible in the figure only 
very close to $\Tc$), compared to the almost saturating behavior in a rather wide temperature range in the 4~Oe case, 
indicates that a finite coercive field immediately emerges just below $\Tc$. However, the coercivity remains small 
(less than 4~Oe) down to about 100~K. 
At this temperature, both the real and imaginary components of the magnetization decrease steeply. 
In comparison, the multilayer samples show a stronger response to a 0.06~Oe ac field, with significant magnitude of both the real and the imaginary parts, well below $\Tc$.  Furthermore, the 4~Oe results show that the decay of the real and imaginary  magnetization components of the multilayers at low temperature is 
more gradual compared to the 300~{\AA}  sample.

\begin{figure}[t]
\includegraphics[width=7.5cm]{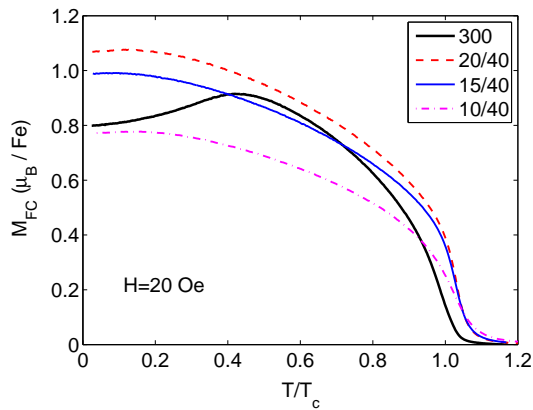}
\centering
\caption{
(Color online) Field-cooled (FC) magnetization measured in a field of 20~Oe versus the reduced temperature $T/\Tc$ for the 
300~\AA (thick solid line), 20/40 (dashed line), 15/40 (solid line) and 10/40 (dashed-dotted line) samples.
\label{FC-reducedT}
}
\end{figure}

The field-induced magnetization above the Curie temperature is enhanced in the multilayers compared to the reference sample. To illustrate this point, we plot the FC magnetization in an applied field of 20~Oe 
in Fig.~\ref{FC-reducedT}, as a function of reduced temperature $T$/$\Tc$. The magnetization shows a sharp increase at a 
temperature above $\Tc$ in a rather similar way for the multilayers, whereas the increase starts much closer to $\Tc$ for the 300~{\AA}
reference sample. This observation is consistent with the previously reported $2D$ $XY$-like character of similar thin amorphous multilayers,\cite{Liebig11,Martina11} resulting in enhanced magnetic correlations above the Curie temperature. \cite{Kosterlitz1974,Bramwell1993}
 The 300~{\AA} {\FeZr} sample has a clear maximum in the FC magnetization at about 0.5 in reduced temperature. 
This reflects an increased importance of exchange frustration and random anisotropy, 
which are responsible for an  increasingly non-collinear arrangement of the magnetic moments with decreasing temperature. 
\cite{Kaul91,Kaul92B,Lorenz95,Uchida01,Park10,Kaul05REV} 
The FC magnetization curves of the multilayer samples also show maxima or plateaus. However, these are much weaker and 
appear at lower temperatures as compared to the 300~{\AA} film. 
This is  in agreement with the suppression of the non-collinearity observed in Fig.~\ref{MH} for the thin {\FeZr} layers.

\begin{figure*}[!]
\includegraphics[width=14cm]{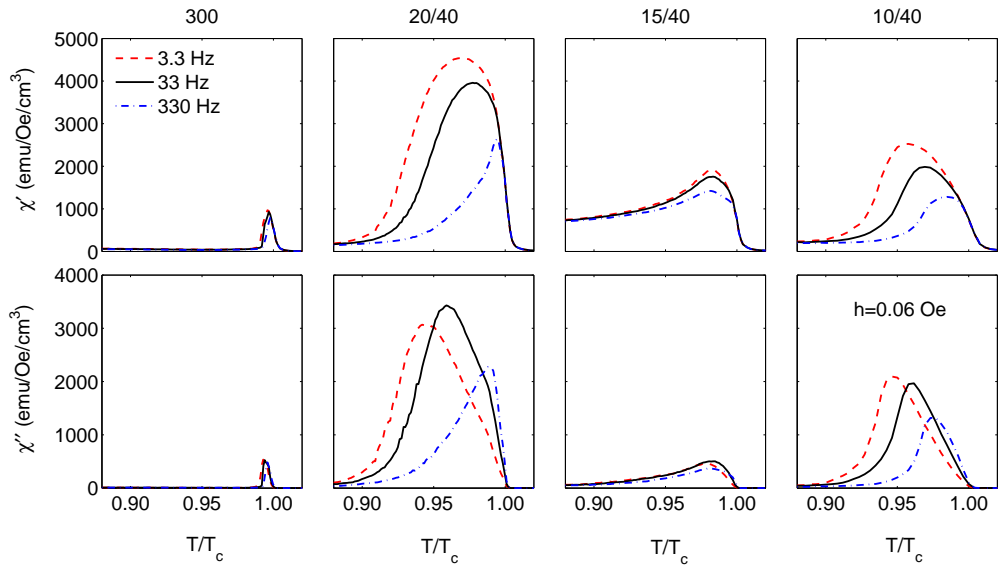}
\centering
\caption{
(Color online) The in-phase ($\chi'$) and out-of-phase ($\chi''$) components of the ac susceptibility, for ac fields with 0.06~Oe 
(rms) amplitude and frequency 3.3, 33 and 330~Hz, plotted versus the reduced temperature ($T/\Tc$).
\label{ac-freq}
}
\end{figure*}

The frequency dependence of the ac susceptibility is displayed in Fig.~\ref{ac-freq}.  
Here, we 
include the ac susceptibility for three different frequencies: 3.3, 33 and 330~Hz; all in an ac field of 0.06~Oe. 
The temperature scale is given in reduced temperature $T/\Tc$, normalized to the $\Tc$ of each sample. 
Starting from high temperature, the real component of the susceptibility $\chi'(T)$ 
is frequency-independent
down to the inflection point (which was used to define $\Tc$) for all samples.
Compared to the reference sample, the multilayers show larger maximum values of $\chi'(T)$ and enhanced regions 
close to $\Tc$, where fluctuations of the magnetization yield strongly frequency-dependent behavior, on the timescale of 
our experiments (1/$\omega$ in the order of milliseconds to seconds). 
Furthermore, the ac susceptibility of the thin {\FeZr} layers remains substantially larger (and frequency dependent) for an 
extended temperature range below $\Tc$, compared to the thick reference film. 
The 20/40 and 10/40 samples behave quite similarly, with the later having a lower magnitude of the susceptibility at all 
frequencies. Compared to them, the susceptibility of the 15/40 sample, which exhibits less perfect  layering, has a smaller amplitude 
and weaker frequency dependence.
The extended range of higher magnetic 
susceptibility for the multilayers is consistent with the presence of a $2D$ $XY$-like magnetization in thin {\FeZr} layers. \cite{Liebig11, Martina11} 
The inherent origin of this enhancement is the wide temperature range of the criticality 
in $2D$ $XY$ systems, as discussed by Archambault et al. [\onlinecite{Archambault97}].

\begin{figure}[b]
\includegraphics[width=7cm]{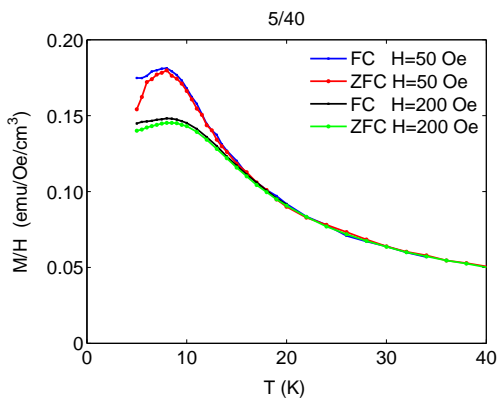}
\centering
\caption{
(Color online) Zero-field cooled (ZFC) and field-cooled (FC) magnetization measured in a field of 50 and 200~Oe versus temperature for the 5/40 multilayer sample.
\label{FAZ5}
}
\end{figure}

We now turn our attention to the Langevin-like $M(H)$ curve of the 5/40 sample (see Fig.~\ref{MH}). A similar behavior has previously been reported for a multilayer of {Co}$_{68}${Fe}$_{24}${Zr}$_{8}$/{Al}$_{2}${O}$_{3}$. \cite{Lidbaum10} However, in
that case, the superparamagnetic behavior could be assigned to the formation of clusters resulting from a collapse of the continuity of the magnetic layers at a thickness of about 10~{\AA}. In the present case, the 5~{\AA} {\FeZr} layers are continuous and do not break down into small structural regions, as confirmed by the X-ray reflectivity measurements (see Fig.~\ref{reflectivity}). 
Therefore, the seemingly superparamagnetic behavior does not arise from geometrically distinguishable regions. 
Long range ferromagnetic order is nevertheless hindered in the 5/40 sample. FC and ZFC magnetization curves measured in applied fields of 50 and 200 Oe are shown in Fig.~\ref{FAZ5}. 
The FC and ZFC curves do not indicate long-range ferromagnetic order at any temperature, but some irreversibility occurs between them below 10 K.
We conclude that only short range magnetically correlated regions form within  the ultrathin {\FeZr} layers.
In order to estimate the size of the correlated magnetic regions, an $M(H)$ curve at 20~K was fitted to a Langevin function, yielding an average magnetic moment of 600~$\mu_B$ for the correlated regions; corresponding to approximately 500 Fe atoms, assuming a moment of 1.3~$\mu_B$ per Fe atom. When considering the thickness of the {\FeZr} bilayer, the in-plane correlation length is of the order of 30~\AA.

\begin{figure}[h]
\includegraphics[width=7.5cm]{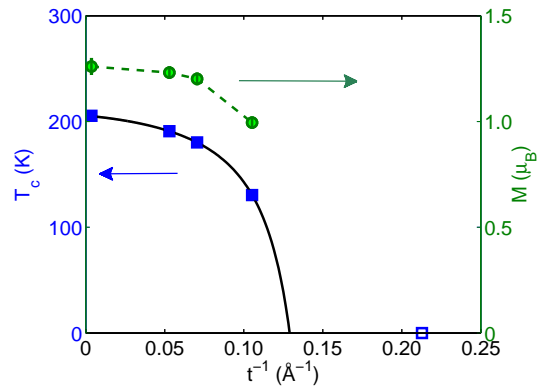}
\centering
\caption{
(Color online) $\Tc$ (squares) and magnetic moment (circles) at 5~K, versus the inverse thickness of the {\FeZr} layers $t^{-1}$. 
The solid line is a fit to the finite-size scaling relation for $\Tc(t)$ [Eq.~(\ref{eq:FSS})] yielding $t'=6.5$~\AA, $t_0=1.2$~\AA, and $\lambda=1.2$.  
For the 300~{\AA} sample, the magnetic moment is determined in a field of 4~T.
\label{muB_Tc}
}
\end{figure}

We will now address the thickness-dependent changes in Curie temperature and discuss the absence of ferromagnetic ordering in the 5/40 sample. 
The thickness dependence of  $\Tc$ is plotted versus the inverse thickness of the {\FeZr} layer in Fig.~\ref{muB_Tc}.  
A fit to Eq.~\ref{eq:FSS} is shown as a solid line in the figure. The best fit was obtained with $t'=6.5$~{\AA} and $\lambda=1/\nu=1.2$. This value of the shift exponent $\lambda$ agrees with those obtained for crystalline films, for which the $\lambda$ values are typically found in between those expected for a 3D Ising system ($1/\nu_{\rm 3D}=1.6$) and for a 2D Ising system ($1/\nu_{\rm 2D}=1$).\cite{Vaz08R}
The 5/40 sample is below the critical thickness $t'$ for a non-zero Curie temperature and the absence of ferromagnetism for the 5/40 sample is hence to be expected. 
The thickness dependence of the magnetic moment in a field of 4~T is also included in Fig.~\ref{muB_Tc}.  The magnetic moment of the multilayer samples is saturated in this field, while that of the  300~{\AA} sample exhibiting a non-collinear magnetization is not fully saturated. The magnetic moment is nevertheless reduced as the {\FeZr} layer thickness decreases.

For ultrathin layers of Fe grown epitaxially, the magnetic moment per Fe atom can be either enhanced or reduced depending on the substrate, e.g. a monolayer of Fe on V(110) is nonmagnetic while a monolayer of Fe on Au(111) exhibits a magnetic moment of 2.9~$\mu_B$.\cite{Nawrath1997,Vaz08R} 
The substrate (or the nonmagnetic spacer in a superlattice) can induce strain on the Fe due to the mismatch in lattice parameter. Furthermore, electronic effects at the interfaces (hybridization effect) between the Fe and substrate/spacer may be an important factor determining the enhancement or reduction of the magnetic moment.
A reduction of the magnetic moment is often described by a magnetically dead layer at each interface in a multilayer structure.
The magnetic moment is, in that case, found to scale with the thickness of Fe as $M(t)/M(\infty) = (1 - 2t_i/t)$ were $t_i$ is the thickness of the interface regions with reduced magnetic moment. \cite{Kim2001}
The reduction of the magnetic moment of {\FeZr} with decreased thickness is, at least to a first approximation, consistent with the magnetically dead layer picture (see Fig.~\ref{muB_Tc}). However, since the saturated moment of the 300~{\AA} sample is not reached in a field of 4~T, a detailed analysis is not possible.
For Fe on V(110) such a magnetically dead layer has been explained by intermixing of Fe and V at the interface. \cite{Izquierdo2001,Yartseva2005}
The interfaces between  amorphous {\FeZr} and {\AlZr} are smooth and suffer the least possible from intermixing due to the low deposition temperatures compared to crystalline films. \cite{Korelis10}
Consequently, the differences
in magnetic properties near the interfaces can be attributed to the change of the chemical environment of the Fe atoms close to interfaces, compared to Fe atoms within the layer. This effect is twofold in {\AlZr/\FeZr} multilayers: First, the effective Zr concentration is higher near the interface to {\AlZr} than within the {\FeZr} layer. Second, the number of nearest Fe-Fe neighbors is lower at the interfaces, reducing the strength of direct exchange coupling between the Fe atoms at the near-interface regions.

\section{Conclusions}

We have investigated how the thickness of the {\FeZr} layers influences 
the magnetic properties of amorphous {\FeZr/\AlZr} multilayers with negligible interlayer coupling.
Both the Curie temperature and magnetic moment are reduced with decreasing  thickness of the {\FeZr} layer. A finite-size scaling of the Curie temperature according to Eq.~(\ref{eq:FSS}) yields the shift exponent $\lambda=1.2$ and the critical thickness $t'=6.5$~{\AA}, below which the Curie temperature is zero. The absence of ferromagnetic order observed in 5~{\AA} {\FeZr} layers is hence in agreement with the finite-size scaling.  
For a 300~{\AA} thick {\FeZr} film, used as bulk reference, the results are consistent with the presence of a non-collinear magnetization. The non-collinear magnetization is greatly reduced for 10-20~{\AA} thick {\FeZr} layers.

\section*{Acknowledgments}

Financial support from the Swedish Research Council (Vetenskapsr\aa det) and the G{\"o}ran Gustafsson Foundation
is gratefully acknowledged.\\


\begin{thebibliography}{10}

\bibitem{Bland94B}
{\em Ultrathin Magnetic Structures I: An Introduction to the Electronic,
  Magnetic and Structural Properties}, edited by J.~A.~C. Bland and B. Heinrich
  (Springer-Verlag, 1994).

\bibitem{Freeman92R}
A. Freeman and R. Wu, J. Magn. Magn. Mater. {\bf 104-107},  1   (1992).

\bibitem{Poulop99R}
P. Poulopoulos and K. Baberschke, J. Phys. Condens. Matter {\bf 11},  9495
  (1999).

\bibitem{Bader94}
S.~D. Bader, D. Li, and Z.~Q. Qiu, J. Appl. Phys. {\bf 76},  6419  (1994).

\bibitem{Vaz08R}
C.~A.~F. Vaz, J.~A.~C. Bland, and G. Lauhoff, Rep. Prog. Phys. {\bf 71},
  056501  (2008).

\bibitem{Allan70}
G.~A.~T. Allan, Phys. Rev. B {\bf 1},  352  (1970).

\bibitem{Fisher72}
M.~E. Fisher and M.~N. Barber, Phys. Rev. Lett. {\bf 28},  1516  (1972).

\bibitem{Li92}
Y. Li and K. Baberschke, Phys. Rev. Lett. {\bf 68},  1208  (1992).

\bibitem{Huang93}
F. Huang, G.~J. Mankey, M.~T. Kief, and R.~F. Willis, J. Appl. Phys. {\bf 73},
  6760  (1993).

\bibitem{Kenning87}
G.~G. Kenning, J.~M. Slaughter, and J.~A. Cowen, Phys. Rev. Lett. {\bf 59},
  2596  (1987).

\bibitem{Thomas92}
A. Thomas and M. Gibbs, J. Magn. Magn. Mater. {\bf 103},  97   (1992).

\bibitem{Suzuki08}
K. Suzuki, N. Ito, J. Garitaonandia, J. Cashion, and G. Herzer, J. Non-Cryst.
  Solids {\bf 354},  5089   (2008).

\bibitem{Castano99}
F.~J. Casta{\~n}o, T. Stobiecki, M.~R.~J. Gibbs, M. Czapkiewicz, J. Wrona, and
  M. Kopcewicz, Thin Solid Films {\bf 348},  233   (1999).

\bibitem{Dubowik95}
J. Dubowik, J. Magn. Magn. Mater. {\bf 140-144},  531   (1995).

\bibitem{Pan99}
F. Pan, M. Zhang, X. tao Liu, and Z. shu Zhang, Jpn. J. Appl. Phys. {\bf 38},
  1383  (1999).

\bibitem{Kopcewicz97}
M. Kopcewicz, T. Stobiecki, M. Czapkiewicz, and A. Grabias, J. Phys. Condens.
  Matter {\bf 9},  103  (1997).

\bibitem{Handschuh93}
S. Handschuh, J. Landes, U. K{\"o}bler, C. Sauer, G. Kisters, A. Fuss, and W.
  Zinn, J. Magn. Magn. Mater. {\bf 119},  254   (1993).

\bibitem{Honda94}
S. Honda and M. Nawate, J. Magn. Magn. Mater. {\bf 136},  163   (1994).

\bibitem{Thiele93}
J. Thiele, F. Klose, A. Schurian, O. Schulte, W. Felsch, and O. Bremert, J.
  Magn. Magn. Mater. {\bf 119},  141   (1993).

\bibitem{Landes91}
J. Landes, C. Sauer, B. Kabius, and W. Zinn, Phys. Rev. B {\bf 44},  8342
  (1991).

\bibitem{Mayr02}
S.~G. Mayr and K. Samwer, Phys. Rev. B {\bf 65},  115408  (2002).

\bibitem{Geisler96}
H. Geisler, U. Herr, T. Lorenz, and K. Samwer, Thin Solid Films {\bf 275},  176
    (1996).

\bibitem{Hiroyoshi82}
H. Hiroyoshi and K. Fukamichi, J. Appl. Phys. {\bf 53},  2226  (1982).

\bibitem{Fukamichi82}
K. Fukamichi, R.~J. Gambino, and T.~R. McGuire, J. Appl. Phys. {\bf 53},  2310
  (1982).

\bibitem{Shirakawa83}
K. Shirakawa, K. Fukamichi, T. Kaneko, and T. Masumoto, Physica B+C {\bf 119},
  192   (1983).

\bibitem{Read84}
D. Read, T. Moyo, and G. Hallam, J. Magn. Magn. Mater. {\bf 44},  279   (1984).

\bibitem{Saito86}
N. Saito, H. Hiroyoshi, K. Fukamichi, and Y. Nakagawa, J. Phys. F: Metal Phys.
  {\bf 16},  911  (1986).

\bibitem{Ryan87}
D.~H. Ryan, J.~M.~D. Coey, E. Batalla, Z. Altounian, and J.~O. Str{\"o}m-Olsen,
  Phys. Rev. B {\bf 35},  8630  (1987).

\bibitem{Tange89}
H. Tange, Y. Tanaka, M. Goto, and K. Fukamichi, J. Magn. Magn. Mater. {\bf 81},
   L243   (1989).

\bibitem{Kaul91}
S.~N. Kaul, J. Phys. Condens. Matter {\bf 3},  4027  (1991).

\bibitem{Kaul92B}
S.~N. Kaul, V. Siruguri, and G. Chandra, Phys. Rev. B {\bf 45},  12343  (1992).

\bibitem{Kiss94}
L.~F. Kiss, T. Kem{\'e}ny, I. Vincze, and L. Gr{\'a}n{\'a}sy, J. Magn. Magn.
  Mater. {\bf 135},  161   (1994).

\bibitem{Ren95}
H. Ren and D.~H. Ryan, Phys. Rev. B {\bf 51},  15885  (1995).

\bibitem{Wildes03}
A.~R. Wildes, J.~R. Stewart, N. Cowlam, S. Al-Heniti, L.~F. Kiss, and T.
  Kem{\'e}ny, J. Phys. Condens. Matter {\bf 15},  675  (2003).

\bibitem{Calderon05}
R.~G. Calderon, L.~F. Barquin, S.~N. Kaul, J.~C.~G. Sal, P. Gorria, J.~S.
  Pedersen, and R.~K. Heenan, Phys. Rev. B {\bf 71},  134413  (2005).

\bibitem{Lorenz95}
R. Lorenz and J. Hafner, J. Magn. Magn. Mater. {\bf 139},  209   (1995).

\bibitem{Uchida01}
T. Uchida and Y. Kakehashi, Phys. Rev. B {\bf 64},  054402  (2001).

\bibitem{Park10}
J.~H. Park and B.~I. Min, Journal of Magnetics {\bf 15},  1  (2010).

\bibitem{Kaul05REV}
S.~N. Kaul, Curr. Sci. {\bf 88},  78  (2005).

\bibitem{Liebig11}
A. Liebig, P.~T. Korelis, M. Ahlberg, and B. Hj{\"o}rvarsson, Phys. Rev. B {\bf
  84},  024430  (2011).

\bibitem{Korelis10}
P.~T. Korelis, A. Liebig, M. Bj{\"o}rck, B. Hj{\"o}rvarsson, H. Lidbaum, K.
  Leifer, and A.~R. Wildes, Thin Solid Films {\bf 519},  404   (2010).

\bibitem{Martina11}
M. Ahlberg, G. Andersson, and B. Hj{\"o}rvarsson, Phys. Rev. B {\bf 83},
  224404  (2011).

\bibitem{Kosterlitz1974}
J.~M. Kosterlitz, J. Phys. C {\bf 7},  1046  (1974).

\bibitem{Bramwell1993}
S.~T. Bramwell and P.~C.~W. Holdsworth, J. Phys. Condens. Matter {\bf 5},  L53
  (1993).

\bibitem{Archambault97}
P. Archambault, S.~T. Bramwell, and P.~C.~W. Holdsworth, J. Phys. A {\bf 30},
  8363  (1997).

\bibitem{Lidbaum10}
H. Lidbaum, H. Raanaei, E.~T. Papaioannou, K. Leifer, and B. Hj{\"o}rvarsson,
  J. Cryst. Growth {\bf 312},  580   (2010).

\bibitem{Nawrath1997}
T. Nawrath, H. Fritzsche, F. Klose, J. Nowikow, C. Polaczyk, and H. Maletta,
  Physica B {\bf 234-236},  505   (1997).

\bibitem{Kim2001}
S.-K. Kim, J.-R. Jeong, J.~B. Kortright, and S.-C. Shin, Phys. Rev. B {\bf 64},
   052406  (2001).

\bibitem{Izquierdo2001}
J. Izquierdo, R. Robles, A. Vega, M. Talanana, and C. Demangeat, Phys. Rev. B
  {\bf 64},  060404  (2001).

\bibitem{Yartseva2005}
N.~S. Yartseva, S.~V. Yartsev, N.~G. Bebenin, and C. Demangeat, Phys. Rev. B
  {\bf 71},  144428  (2005).

\end{thebibliography}
\end{document}